\documentclass[referee]{raa}           
\usepackage{graphicx,times}
\usepackage{natbib}
\usepackage{amssymb,amsmath}
\usepackage{amsmath}
\usepackage{soul}
\usepackage{color}
\bibpunct{(}{)}{;}{a}{}{,}

\usepackage[a4paper=true,pagebackref=true]{hyperref}
\hypersetup{pdftitle = The title of my PDF, pdfauthor = My name, pdfsubject= The subject, pdfkeywords = keyword1 keyword2 keyword3} 
\hypersetup{colorlinks = true, linkcolor = green, anchorcolor = red, citecolor = blue, filecolor = red, pagecolor = red, urlcolor = red}

\begin{document}

   \title{The unusual emission from PSR B1859$+$07 with FAST
}

 \volnopage{ {\bf 2021} Vol.\ {\bf X} No. {\bf XX}, 000--000}
   \setcounter{page}{1}

 \author{Lin Wang\inst{1,3}, Ye-Zhao Yu\inst{2,1}, Feifei Kou\inst{5}, Kuo Liu\inst{4},  Xinxin Wang\inst{6}, Bo Peng\inst{1}}

   \institute{CAS Key Laboratory of FAST, National Astronomical Observatories, Chinese Academy of Sciences,
             Beijing, 100101, P. R. China; {\it pb@bao.ac.cn}\\
        \and
             Qiannan Normal University for Nationalities,
             Duyun, 558000, P. R. China; {\it yuyezhao@foxmail.com}\\
    \and
    School of Astronomy and Space Science, University of Chinese Academy of Sciences, Beijing, 100049, P. R. China\\
    \and
    Max-Plank-Institut f{\"u}r Radioastronomie, Auf dem H{\"u}gel 69, Bonn, D-53121 , Germany\\
    \and 
    Xinjiang Astronomical Observatories, Chinese Academy of Sciences, Urumqi, 830011, P. R. China\\
    \and
    United World College of Changshu China, Changshu, 215500, P. R. China
\vs \no
}

\abstract{We present simultaneous broad-band radio observations on the abnormal emission mode from PSR B1859$+$07 using the Five-hundred-meter Aperture Spherical radio Telescope (FAST). This pulsar shows peculiar emission phenomena, which are occasional shifts of emission to an early rotational phase and mode change of emission at the normal phase. We confirm all these three emission modes with our datasets, including the B (burst) and Q (quiet) modes of the non-shift pulses and the emission shift mode with a quasi-periodicity of 155 pulses. We also identify a new type of emission shift event, which has emission at the normal phase during the event. We studied polarisation properties of these emission modes in details, and found that they all have similar polarisation angle (PA) curve, indicating the emission of all these three modes are from the same emission height. 
}

   \authorrunning{L. Wang et al. }            
   \titlerunning{Unusual emission from PSR B1859$+$07}  
   \maketitle

%
\section{Introduction}           
\label{sect:intro}

The study of pulsar emission is one of the most important subjects in pulsar astronomy, for the main purpose of understanding the underlying physics of the emission process. This has been carried out extensively based on some peculiar emission phenomena such as sub-pulse drifting, nulling, mode changing and so forth. Some of these phenomena have been seen and studied for years and can now be partially explained by theories of pulsar emission. However, some phenomena are still poorly understood, which 
requires more observational facts to build a theoretical system that is consistent with the observation. A remarkable phenomenon of such is called `emission shift', `swoosh' or `flare` in the literature \citep[e.g.][]{2006MNRAS.370..673R,2016MNRAS.461.3740W,2016Perera,2016Han}, firstly identified by \cite{2006MNRAS.370..673R} in PSRs B0919$+$06 and B1859$+$07. This phenomenon is seen as emission from subsequent rotations gradually arriving in earlier pulse phase, then lasting for tens of pulse periods and finally gradually returning back to the usual emission pulse phase. A quasi-periodicity of the emission shift behavior was discovered in PSR B1859+07 and very likely in PSR B0919$+$06 by \cite{2016MNRAS.461.3740W}. Since then, some hypotheses have been raised to explain this peculiar emission phenomenon in these two pulsars \citep{2016MNRAS.461.3740W,2017MNRAS.469.2049Y,2018ApJ...855...35G,2021MNRAS.506.5836R}. \cite{2019SCPMA..6259504Y} studied the emission shift events of PSR B0919$+$06 using multi-frequency FAST (Five-hundred-meter Aperture Spherical radio Telescope) datasets in detail, and found that there might be correlations between this peculiar emission behavior and the pulsar profile changing to some extent. However, the underlying physics of this phenomenon has yet to be revealed.\\

PSR B1859$+$07 has a spin period of 0.64 seconds and a high DM of 252.8\,cm$^{-3}$pc, and is known for its abnormal emission mode as mentioned above. In addition to the emission shift, it has two slightly different profile shapes in its normal emission state. \citet{2016Perera} reported its profile modulation corresponds to a quasi-periodic variation of spin frequency derivative with a period of 350 days. It is similar to PSR B0919$+$06 that two emission states in the normal phase correlate to a quasi-periodically spin frequency switching \citep{2010SLyne}. However, the emission shift states of both PSR B1859$+$07 and PSR B0919$+$06 are not correlated to their spin frequency derivatives. Further studies of the emission shift event are then necessary to investigate whether it is linked to any characteristics of the pulsar.

Taking the advantage of the excellent performance of FAST on individual pulse study for radio pulsars \citep{2019Whg}, we studied emission states of PSR B1859$+$07. In Section \ref{sec:obs}, we introduce the details of our observation and the data analysis of this work. We present our results, including the emission shift state, profile modulation and polarisation in Section \ref{sec:result}. We further discuss the possible correlations between our results and the pulsar characteristics in Section \ref{sec:diss} and summarize the conclusions of this work in Section \ref{sec:summary}.

\section{Observations and data reduction}
\label{sec:obs}

We observed PSR B1859$+$07 with the FAST at multi-frequency since January 2018. Before May 2018, an Ultra-Wideband receiver (UWB) covering frequency range of 270--1620\,MHz \citep{2019SCPMA..6259502J} was used, and we observed PSR B1859$+$07 twice using the UWB. The signal from the UWB was originally filtered into two subbands, 270--1000\,MHz (the low-band) and 1000--1620\,MHz (the high-band), to minimize the influence of strong interference \citep{2020Zhy}.
Both of the two subbands were then sent to a Reconfigurable Open Architecture Computing Hardware 2 (ROACH 2) \footnote{https://casper.berkeley.edu/} unit, and were sampled at Nyquist frequency, digitized into 8-bit real samples with dual polarization, packetized and stored on a hard-drive device in PSRFITS format. The sampling time and channel bandwidth of the UWB data are 100\,$\mu$s and 0.25\,MHz, respectively.\\

After May 2018, the UWB was replaced by a 19-beam receiver (19-beam) with frequency spanning 1050--1450\,MHz \citep{2020RAA....20...64J}. We observed PSR B1859+07 using 19-beam receiver on 28th August 2018, 3rd December 2019 and 22nd November 2020, respectively. The 19-beam data were also sampled and digitized using ROACH unit, then output to 8-bit PSRFITS files. The observation made in August 2018 were recorded with only one polarization due to hardware issue, and the other two datasets were output with full polarization. The time resolution and channel bandwidth of the datasets observed on August 2018 and December 2019 were 49.152\,$\mu$s and 122\,kHz, respectively. The last 19-beam data were sampled with time resolution of 196.608\,$\mu$s and channelized with channel bandwidth of 122\,kHz. Before the observations made both in Dec. 2019 and Nov. 2020, the telescope pointed to a off-source region and tracked for 2 minutes. During the off-source observations, a noise diode with a period of 0.2 second and temperature of 10 Kelvin was injected. The off-source data were recorded with the same time and frequency resolution. A summary of all these observations are listed in Table \ref{tab:obs}.

All five data sets were dedispersed at the best DM of PSR B1859$+$07 and folded using the pulsar software package \textsc{DSPSR} \citep{2011PASA...28....1V}. Each rotation was sampled using 1024 phase bins. After dedispersing and folding, we divided the low-band UWB data into 10 subbands and each has bandwidth of 50\,MHz between 280--780\,MHz. The subbands between 280--380\,MHz  were removed from this work due to strong radio interference. The high-band UWB data were analyzed as a whole between 1250--1550\,MHz, as most of the data were contaminated by radio interference. The 19-beam observations made in Dec. 2019 and Nov. 2020 were polarisation calibrated using the pulsar software package \textsc{PSRCHIVE} \citep{2004PASA...21..302H} after dedispersing and folding. In order to reduce the computing time, we summed every 8 frequency channel, resulting a final channel bandwidth of these two data sets to be 976\,MHz. No polarisation calibration was operated on the data set from Aug. 2018 due to the hardware issue. We then searched for the emission shift events by examining the energy in the pulse region of $-$20$^\circ$ $\sim$ $-$10$^\circ$ (Figure \ref{fig:uwb_sp}) among all these individual pulses. We identified the emission shift event as the significance of the energy in the region of $-$20$^\circ$ $\sim$ $-$10$^\circ$ when being greater than 7-$\sigma$.

\begin{table}
\begin{center}
\caption[]{Observational parameters. In the polarization column, the numbers refer to  1: one of the two linear polarization channels; 2: two linear polarization channels; 4: full polarizations. Events is the number of emission shifting events captured in the data set. UWB is the ultra-wide bandwidth receiver and 19-beam refers to the 19-beam receiver.}\label{tab:obs}
 \begin{tabular}{lcccccccc}
  \hline\noalign{\smallskip}
  Date   & MJD   & Receiver   & Pols & Length   & Events & shift & B mode & Q mode\\
  (year-month-Day) & day &  &   & (pulses) &  &(pulses) &(pulses) &(pulses) \\
  \hline\noalign{\smallskip}
2018-01-23 & 58141  & UWB     & 2  & 2773  & 13 & 411  & 1594 & 768\\ 
2018-02-09 & 58159  & UWB     & 2  & 2811  & 15 & 421  & 1635 & 755\\
2018-08-28 & 58358  & 19-beam & 1  & 2795  & 12 & 466  & 1537 & 792\\
2019-12-03 & 58820  & 19-beam & 4  & 5590  & 29 & 850  & 1543 & 3197\\
2020-11-22 & 59175  & 19-beam & 4  & 8800  & 55 & 1541 & 1980 & 5279\\
  \noalign{\smallskip}\hline
\end{tabular}
\end{center}
\end{table}

\section{Results}
\label{sec:result}

\subsection{emission shift events}

\begin{figure}
   \centering
   \includegraphics[width=\textwidth,angle=0]{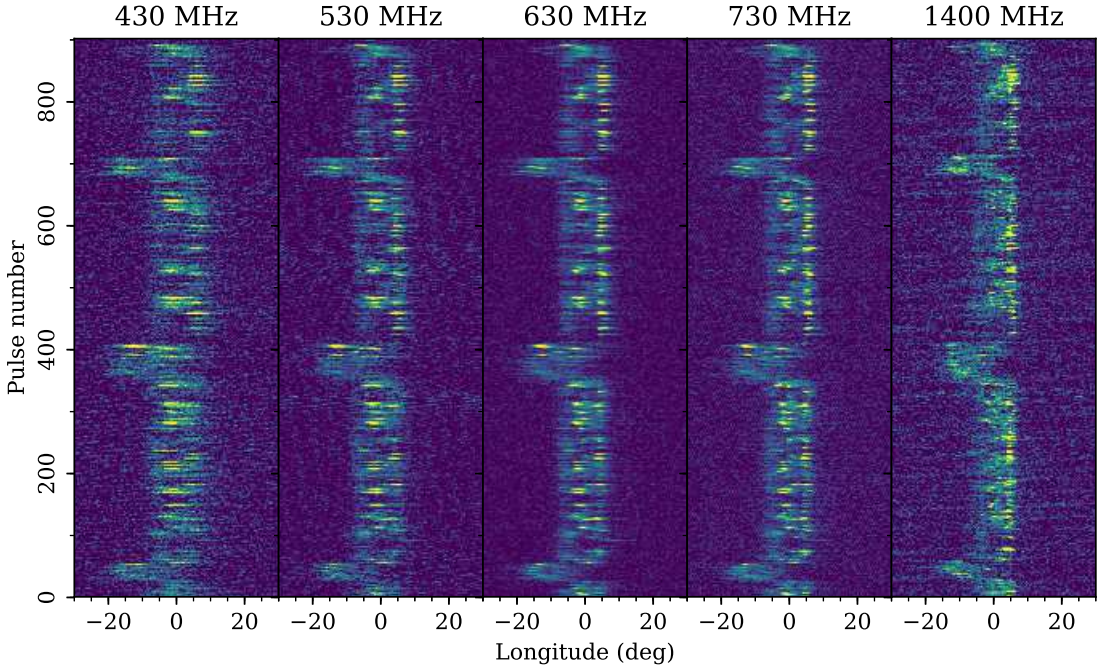}
   \caption{The pulse-stack of PSR B1859$+$07 simultaneously observed with FAST using UWB receiver in February 2018. The panels from left to right are data at centre frequency of 430, 530, 630, 730 and 1400 MHz. The bandwidth of the first 4 panels is 100 MHz and of the fifth panel is 300 MHz. Three emission shift events are displayed at each frequency.}
   \label{fig:uwb_sp}
   \end{figure}

\begin{figure}
   \centering
   \includegraphics[width=\textwidth, angle=0]{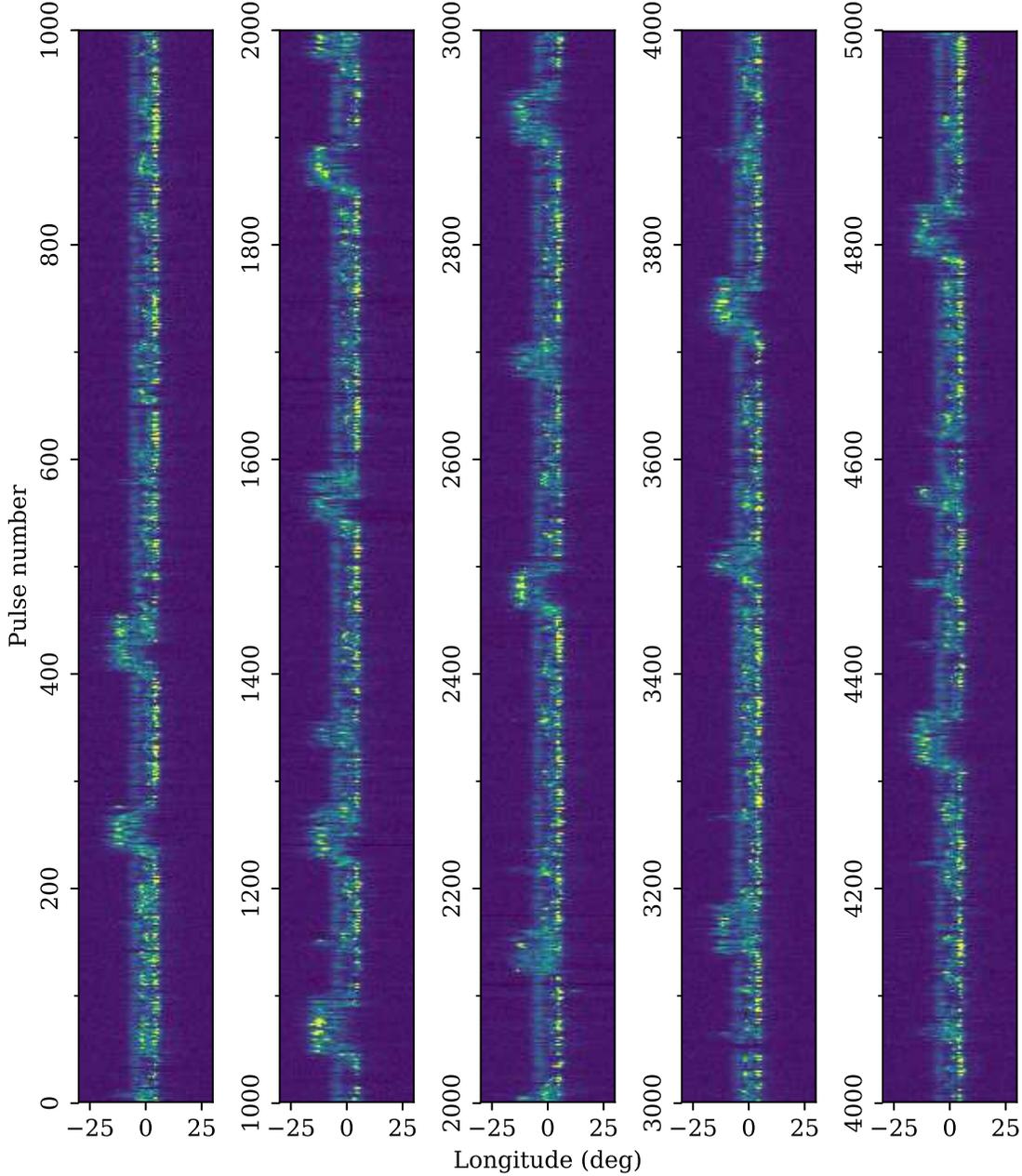}
   \caption{The pulse-stack of B1859$+$07 observed with FAST using 19-beam receiver in November 2020. Both long and short duration emission shift events are displayed.}
   \label{fig:19beam_sp}
   \end{figure}

\begin{figure}
   \centering
   \includegraphics[width=0.5\textwidth, angle=0]{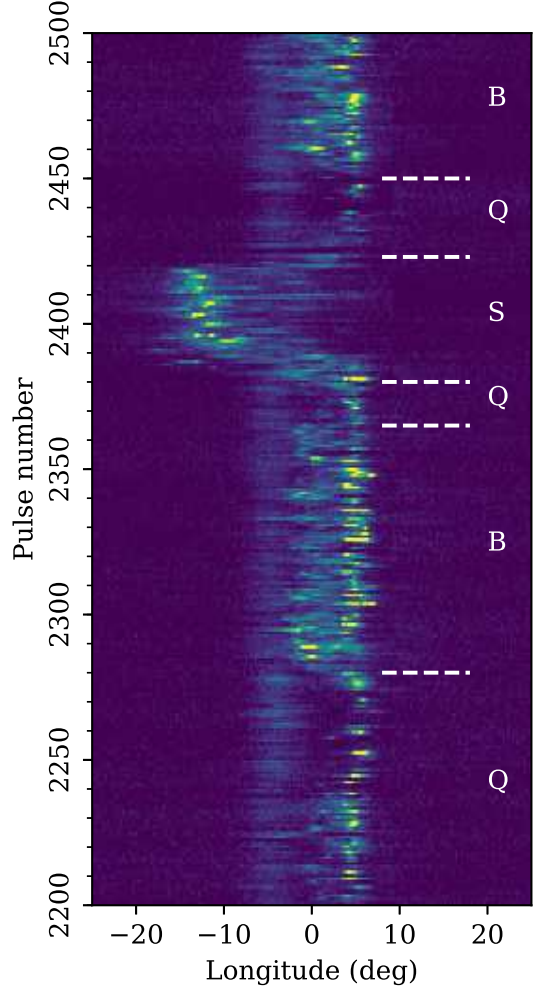}
   \caption{A zoom-in example for the emission shift event of B1859$+$07 observed with FAST using the UWB receiver in August 2018. The B and Q modes are marked with `B' and `Q', respectively. `S' refers to the emission shift event.}
   \label{fig:shift}
   \end{figure}

\begin{figure}
   \centering
   \includegraphics[width=0.5\textwidth, angle=0]{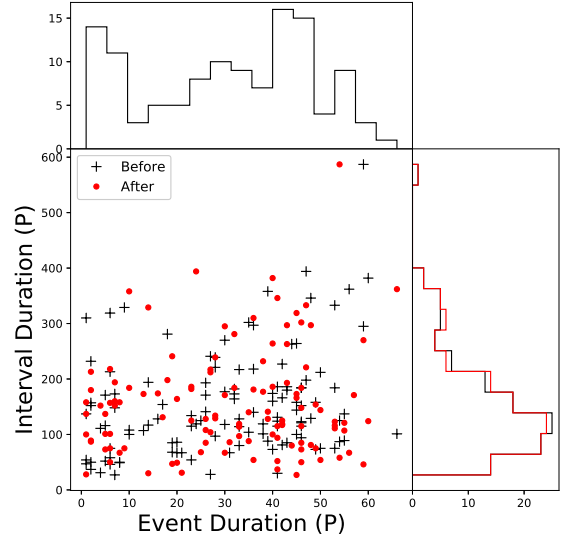}
   \caption{Distribution of the event duration and their interval duration for the 120 emission shift events. The interval duration refers to the time between two adjacent events. In the lower-left panel, the red points are the events with interval duration relative to their former events, and the black crosses are the events with interval duration relative to their later events. The top panel is the distribution of the event duration. The lower-right panel is the distribution of the event interval duration. 30 events are with duration shorter than 15 periods, while 21 of them are captured in the data set observed on 22nd November 2020.}
   \label{fig:event}
   \end{figure}

\begin{figure}
   \centering
   \includegraphics[width=\textwidth, angle=0]{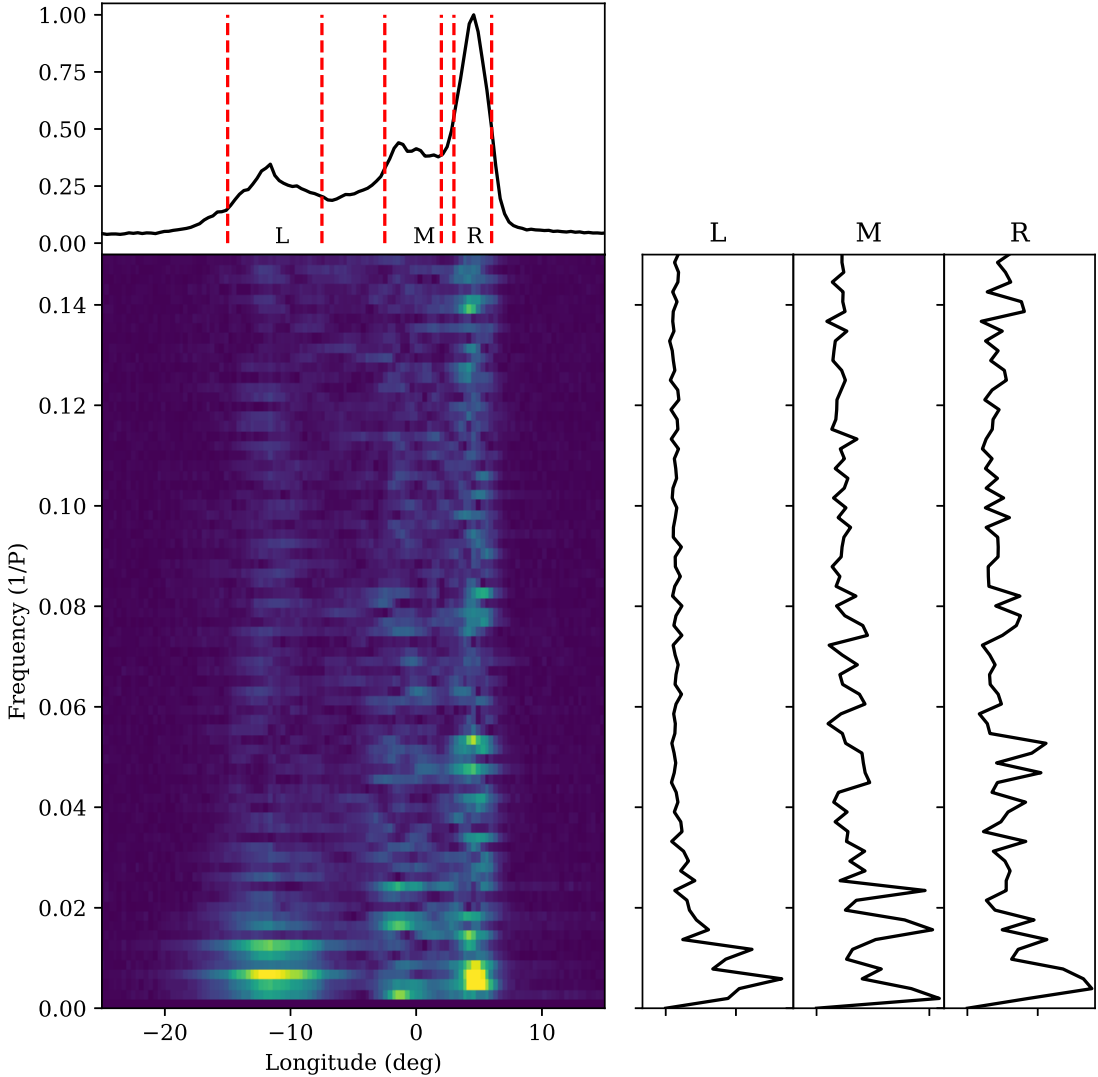}
   \caption{The 2-D fluctuation spectrum of B1859$+$07 observed using FAST on 22nd November 2020.  The summed spectra of L, M and R are shown in the right panels respectively. These summed spectra indicate the L, M and R regions have low frequency peak $\sim$0.006, $\sim$0.016 and $\sim$0.006 cycles per period, which indicate quasi-periods of $\sim$160, $\sim$60 and $\sim$160 pulsar period respectively.}
   \label{fig:FFT}
   \end{figure}

\begin{figure}
   \centering
   \includegraphics[width=0.5\textwidth, angle=0]{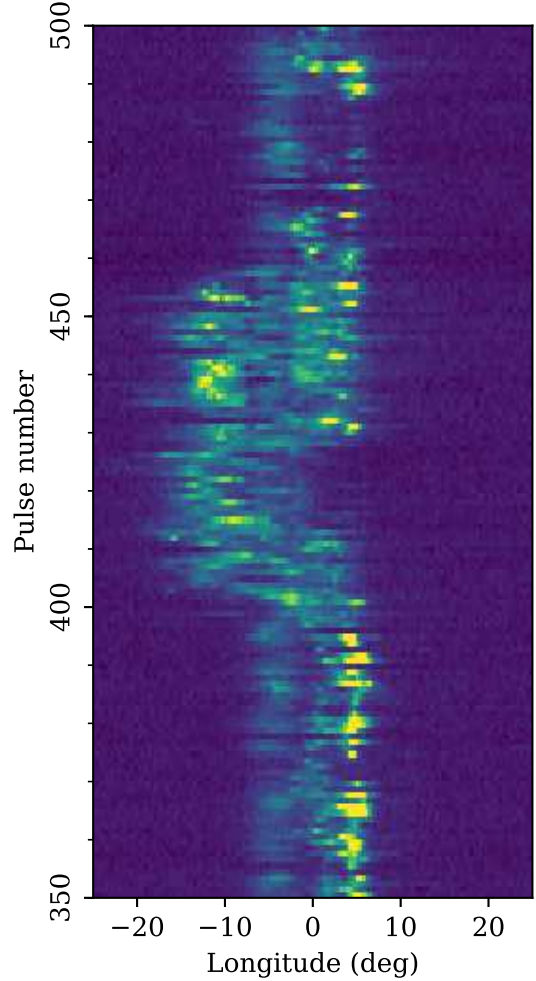}
   \caption{A zoom-in example for the emission shift event of B1859$+$07 observed with FAST using the 19-beam receiver in November 2020.}
   \label{fig:special_shift}
   \end{figure}

\begin{figure}
   \centering
   \includegraphics[width=0.5\textwidth, angle=0]{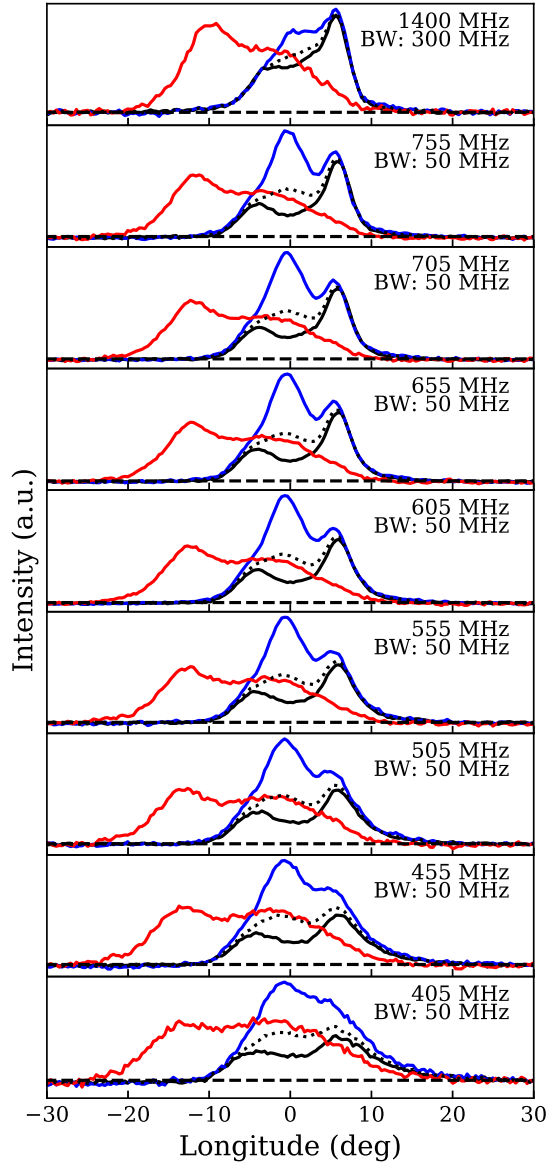}
   \caption{Normalized average profile at different frequencies of each emission state, including emission shift, B mode and Q mode. The data was observed on 2018-02-09 using the UWB receiver. The red solid and the black dotted curves represent the average profiles of the emission shifted and non-shifted pulses, respectively. Black solid and blue curves are the profiles of Q and B modes, respectively among the non-shifted pulses. }
   \label{fig:ave}
   \end{figure}

\begin{figure}
   \centering
   \includegraphics[width=0.5\textwidth, angle=0]{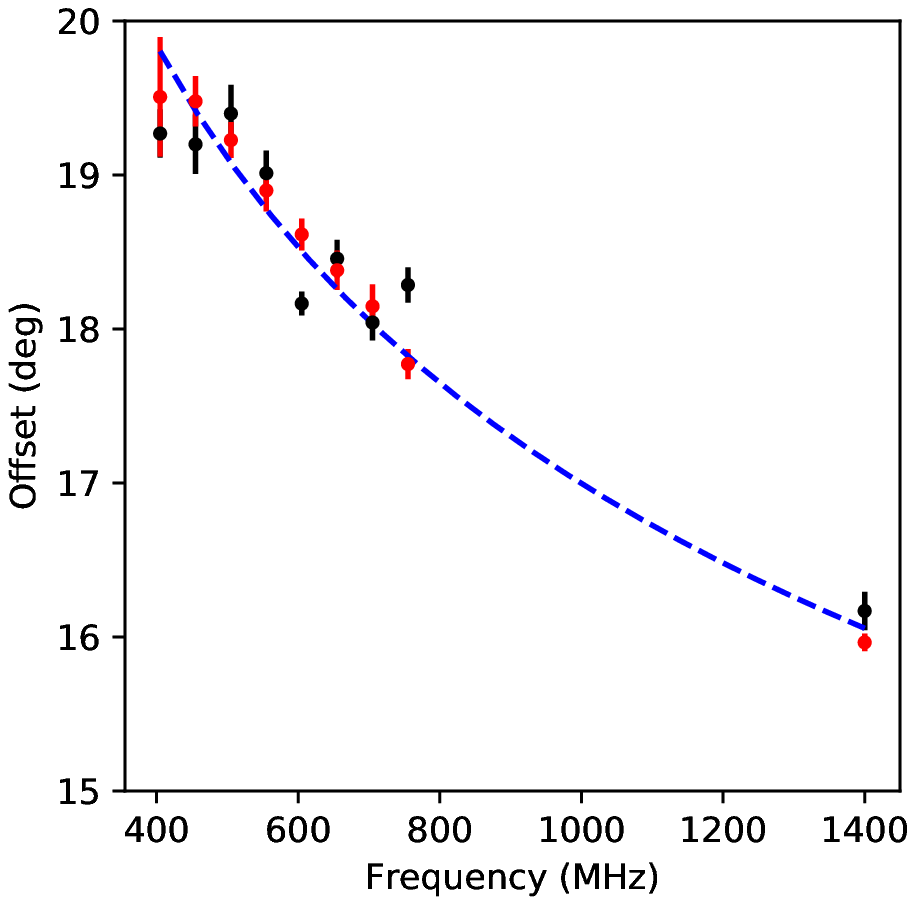}
   \caption{The offset between the non-shift pulses profile and the shift pulses profile as a function of observing frequency for PSR B1859$+$07. Each component is modeled by a Gaussian function and the offset is measured from the peak of shift pulses profile to the non-shift pulses profile. Black and red color corresponds to the data set observed on 2018-01-23 and 2018-02-09, respectively. All data points are with  1-$\sigma$ error bars. Blue dashed curve is the power-law fitting result.}
   \label{fig:power_law}
   \end{figure}

We generated 22$,$769 individual pulses in total, and captured 124 emission shift events. Figure \ref{fig:uwb_sp}  and \ref{fig:19beam_sp} demonstrate examples of the emission shift events captured in the data sets observed in Feb. 2018 and Nov. 2020 at different frequencies, respectively. The former data set is from the UWB receiver and the later are from the 19-beam receiver. Figure \ref{fig:shift} shows the details of one of the emission shift events. The emission from subsequent rotations gradually arrives in earlier pulse phase, then lasts for tens of pulse periods and finally gradually returns back to the normal emission pulse phase.  We analyzed the events duration and the time interval between two adjacent events in Figure \ref{fig:event}.
The duration of most of the emission shift events are longer than 15 pulse periods with a median value of 50 pulse periods approximately. Only 30 shift events are shorter than 15 pulse periods, and 21 out of 30 are from the data set observed in Nov. 2020. We also noticed a median value of the time interval of 155 pulse periods. Using longitude-resolved fluctuation spectrum analysis, we found a quasi-periodicity of 160 pulse periods in the occurrence of these events as shown in Figure \ref{fig:FFT}. This agrees to the median of the time interval of the events, and is consistent with the 150 periods periodicity reported in \citet{2016MNRAS.461.3740W}. Zooming into each emission shift event, we noticed a distinct type of these events. An example of such is displayed in Figure \ref{fig:special_shift}, where emission is present in both shift and non-shift pulse regions, and almost one third of the shift events show behaviour like this.
The emission shift events appear simultaneously in all frequency bands of our data sets (See Figure \ref{fig:uwb_sp}). They are frequency-dependent (Figure \ref{fig:ave} and \ref{fig:power_law}), though this phenomenon is not significant in the individual pulses (Figure \ref{fig:uwb_sp}  and \ref{fig:19beam_sp}). We present the average profile of the shift pulses in Figure \ref{fig:ave}, which shows complex components. The leading component is brighter at high frequency bands, the intensity difference between components becomes less significant as observing frequency decreases. The offsets of the shifted pulse profiles to the non-shift ones show frequency-dependence. This frequency-dependence is contrary to that of PSR B0919+06, that the maximum offset of shifted pulses increases with deceasing frequency \citep{2021MNRAS.506.5836R}. 
In order to model this frequency-dependence, we performed three-component fitting on these profiles, and measured the maximum offset between the peak of the shifted pulse profile and the non-shifted one in each frequency band. Then a power-law is used to model the frequency-dependence of the offset \citep{1975Ruderman}

\begin{equation}
\small
\Delta\theta = A\nu^{-\alpha}
\label{eq:power_law}
\end{equation}

\noindent where $\alpha$ is the separation index. We derived the power-law index $\alpha$ $\sim$ 0.169(8) (See Figure \ref{fig:power_law}).

\subsection{non-shift pulse behaviour}

 As shown in Figure \ref{fig:shift}, the intensity around the longitude of ``0$^\circ$'' fluctuates between a strong and weak state. These are two emission states within the group of non-shift pulses. In order to distinguish these two states, we firstly generated a profile by averaging all the non-shifted pulses. A noise-free template was created by fitting von Mises functions to the integrating average profile. Then we selected the region between longitude $-$2.5$^\circ$ and $+$2.5$^\circ$, and compared the energy of all non-shifted pulses with the template in this region. The pulses that were brighter than the template in this region were defined as B (Burst) mode, otherwise the pulse would be classified as Q (Quiet) mode. Examples of these two modes are marked in Figure \ref{fig:shift}. An intriguing phenomenon is that almost all of the shift events are followed by a series of Q mode pulses in the data sets before 2020, while only half of shift events are seen to be followed by Q mode subsequent pulses in the data set observed in Nov 2020. We integrated the pulses of the B and Q modes in each frequency band and generated average profiles of these two modes in Figure \ref{fig:ave}, respectively. Both profiles of the two modes exhibit complex components, and each component shows evolution with frequencies. In the data sets with centre frequency between 405 $\sim$ 755\,MHz, the leading component of B mode is brighter than its trailing component, and the difference of intensity between the two components decreases as observing frequency decreases. The B mode shows a brighter tailing component at 1.4\,GHz. The Q mode behaves a brighter trailing component in all frequency bands, and the intensity difference between components decreases as decreasing frequency.

\subsection{polarization}

\begin{figure}
   \centering
   \includegraphics[width=0.5\textwidth, angle=0]{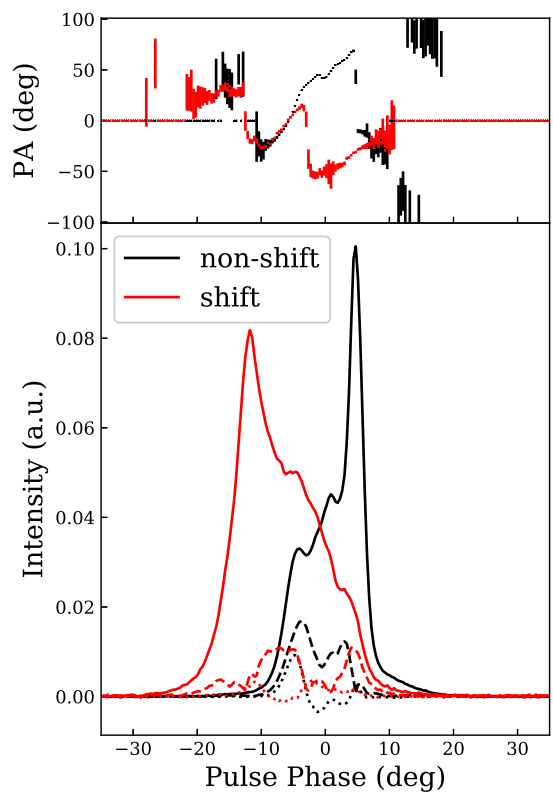}
   \caption{Polarization averaged profiles of B1859$+$07. The data used here are from the observation using the 19-beam receiver on 3rd December 2019. The upper panel shows the curve of the linear polarization angle (PA), $\Phi$. The PA of emission shift and non-shift pulses are marked in red and black respectively. In the bottom panel, red curves are averaged profiles for the emission shift pulses, and black curves are for the non-shift pulses. The solid, dashed and dotted  curves corresponds to total \textit{I}, linear polarization \textit{L} and circular polarization \textit{V} intensity, respectively. The intensity is in arbitrary unit (a.u.). }
   \label{fig:pol_NA}
   \end{figure}

\begin{figure}
   \centering
   \includegraphics[width=0.5\textwidth, angle=0]{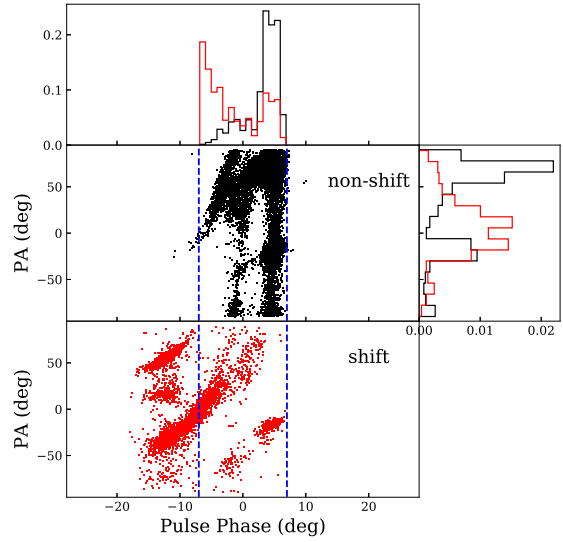}
   \caption{Polarization statistic of the shift and non-shift pulses of B1859$+$07. The data set comes from the observation using 19-beam receiver on 3rd December 2019.  Red and black color represents the PA of shift and non-shift pulses, respectively. The top-left and right panels are the histograms of the normalised probability density distribution. }
   \label{fig:pol_s_NA_19}
   \end{figure}

\begin{figure}
   \centering
   \includegraphics[width=0.5\textwidth, angle=0]{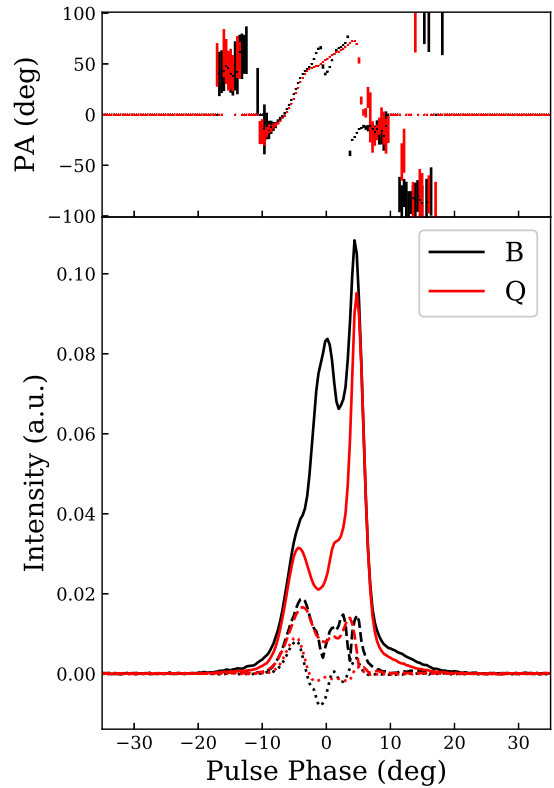}
   \caption{Polarization averaged profiles of PSR B1859$+$07. This data set comes from the observation on 3rd December 2019. Black and red color corresponds to B and Q modes, respectively. In the lower panel, the solid, dashed and dotted curves are total \textit{I}, linear polarization \textit{L} and circular polarization \textit{V} intensity. The upper panel shows the linear polarization angle (PA) curves.}
   \label{fig:pol_BQ_19}
   \end{figure}

Similar to majority pulsars, the polarimetry of average profile in PSR B1859$+$07 is highly linear polarized as demonstrated in Figure \ref{fig:pol_NA}. Similar percentages of linear and circular polarizations are shown in profiles of both shift and non-shift pulses. `Jumps' appear in the Polarization Angle (PA) curves of both shift and non-shift pulse profiles at slightly different longitudes, as demonstrated in Figure \ref{fig:pol_NA}. Since the non-shift pulses dominate between longitude of $-$7$^\circ$ $\sim$ 7$^\circ$, we plot the PA distributions of both shift and non-shift individual pulses in Figure \ref{fig:pol_s_NA_19}. Two orthogonal polarization modes appear simultaneously in the non-shift individual pulses at both sides of longitude 0$^\circ$. For the non-shift individual pulses, their orthogonal polarization modes are present simultaneously at both sides of longitude 0$^\circ$. The dominated polarization mode switches one over the other at the longitude around $8^\circ$, which results in the PA shift for the mean pulse profile (Figure \ref{fig:pol_NA}). The polarized profiles of B and Q modes are displayed in Figure \ref{fig:pol_BQ_19}. The two emission modes have similar polarization properties, although the PA curve of B mode shows fluctuations around longitude of 0$^\circ$. As done for shift and non-shift pulses, we analyzed PA distribution of the B and Q modes. As displayed in Figure \ref{fig:pol_s_NA_19}, the two modes have very similar PA curve, except that there are few Q mode pulses with PA around $-$40$^\circ$. 


   
   \begin{figure}
   \centering
   \includegraphics[width=0.5\textwidth, angle=0]{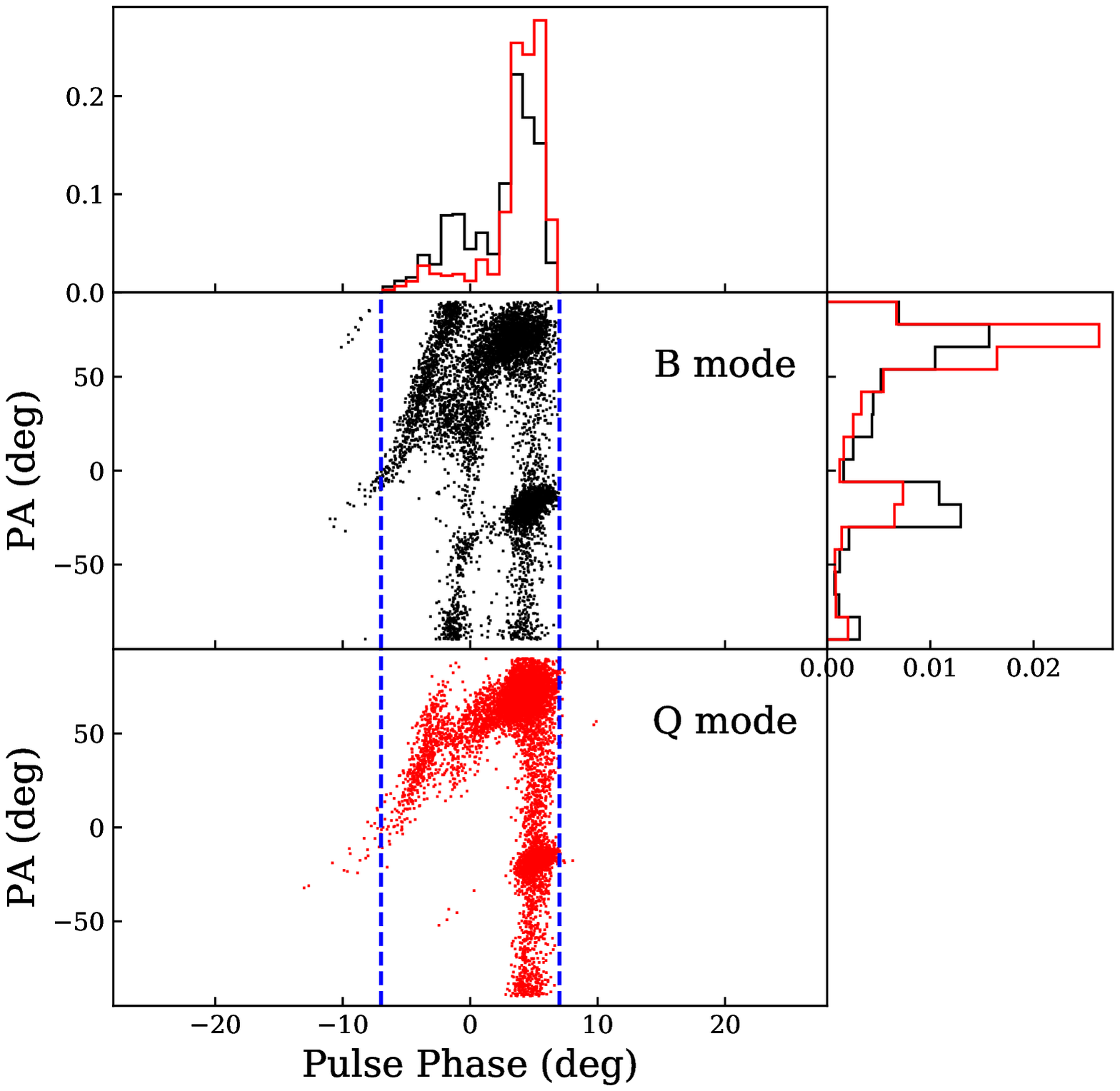}
   \caption{Polarization statistic of the B and Q modes of B1859$+$07. The data set comes form the observation using 19-beam receiver on 3rd December 2019.  Red and black color represents the PA of Q and B modes, respectively. The top-left and right panels are the histograms of the normalized probability density distributions. }
   \label{pol_s_BQ_19}
   \end{figure}

\section{Discussion }
\label{sec:diss}

Tracing the emission and propagation of the pulsar signal, the emission shift events in principle could be resultant of asymmetric of pulsar emission beam, the change of pulsar magnetosphere or dynamic effects around the pulsar (e.g. tidal effects from an ultra-compact binary). \citet{2016MNRAS.461.3740W} and \citet{2018ApJ...855...35G} suggested that the emission shift events may reflect existence of an orbiting body. While we suspect that this is unlikely the reason based on the work done on this pulsar in the literature. If we assume this pulsar is a member of an ultra-compact system and the emission shift events are due to the tidal effect of the companion, the observed spin period derivative would be affected by the orbit motion and thus correlated to the events. However, \citet{2016Perera} investigated the link between the spin period derivative and the emission shift events, no correlation was found therein. Thus the emission shift event in PSR B1859+07 are not likely linked to the orbit motion.

Based on radius-to-frequency mapping (RFM; \citet{1978Cordes}) in pulsars, the pulse profile is narrower at high frequency than that at low frequency. Earlier observations confirmed that the separation between components of pulsar profile decrease with increasing frequency. The frequency dependence of the components separation can be described by Equation \ref{eq:power_law}. \citet{1977M&T} modeled this relationship for several pulsars, and found  0.08 $\leq$ $\alpha$ $\leq$ 0.5 when frequency is below 1.4\,GHz. Our results reveal that the offset of the shifted pulse decreases with increasing frequency following the relationship of Equation \ref{eq:power_law}, and derived the power-law index $\alpha$ $\sim$ 0.169(8) (See Figure \ref{fig:power_law}). Intriguingly, PSR B0919$+$06 is another known pulsar for its emission shift events, but shifted offset increases with increasing frequency \citep{2021MNRAS.506.5836R}. Obviously, the events from the two pulsars can not be described simply using RFM. If we attribute the emission shift events of both PSRs B0919+06 and B1859+07 to different origins, the emission of PSR B1859+07 tends to be under the frame of RFM since the power-law index in Equation \ref{eq:power_law} from our result is in agreement with \citet{1977M&T}. If the emission shift events of the two pulsars have the same origin, it is likely that the shift and non-shift components have different radiation centers, thus Equation \ref{eq:power_law} is not applicable to describe the relationship between component offsets and frequencies.  \citet{2006MNRAS.370..673R} explained the behaviour of these two pulsars based on the core-cone model, claiming that the usual emission of these pulsars were due to a partially active cone – the leading part of the cone is inactive and active during the non-shift and shift emission modes respectively while the trailing part is inverse. This scenario is not able to explain all the behaviours of these pulsars, such as the duration of the emission shift events. An alternative model was put forward to explain the emission features of these pulsars by  \citet{2021MNRAS.506.5836R}, based on the shrinking and expanding  magnetosphere of the neutron star proposed in \citet{2010Timkhin}. \citet{2021MNRAS.506.5836R} explains the emission beam to be a patchy fan beam, and the polar cap region expands as the magnetosphere shrinking and thus changing the angle between the emission beaming direction and our LOS (line of sight). Emissions from different heights may stay in or out of LOS depending on the scale of shrinking (See Figure 10 in \citet{2021MNRAS.506.5836R}).  A simple way to test whether this scenario is applicable is to measure the emission heights of the three emission states. While it is not easy to measure the emission heights, we can still parse this problem in a qualitative way. The PA reflects the direction of the magnetic field at the point of emission and PAs at the same pulse phase are in principle different due to aberration, retardation and field line distortion of a pulsar. Our results show that all the three emission modes have similar PA distributions, thus we suspect the emission at the same phase corresponds to similar emission height. The three modes are unlikely to be resultant of the propagation effect which is a gradual process in usual sense. The emission shift events with emission at both the shift and non-shift pulse regions are also difficult to be explained by the model in \citet{2021MNRAS.506.5836R}.

We noticed that component L and R have the same modulation periods of $\sim$ 160 pulse periods, while the component M are with the modulation period of 60 spin periods in Figure \ref{fig:FFT}, the component L refers to the emission shift events. When the shift events occur, the source is usually bright in component L but faint in component R, which results in the same cadence of flux variation and thus leads to the same modulation period of L and R. The component M corresponds to the mode changing (B and Q modes) state, and no evidence shows that the mode changing correlates to the emission shift events. Considering the different modulation periods between components M and the other components (L and R), it is likely that component M is originated from different regions to components L and R. However, whether these emission modes are due to intrinsic change of radiative particles from the pulsar would be an interesting but challenging question. Further observations  using high sensitivity telescopes, especially FAST on a larger sample of this kind of pulsar would facilitate the solution of this question.


\section{Summary}
\label{sec:summary}

We present FAST observations of PSR B1859$+$07 at 0.4 $\sim$ 1400\,MHz using UWB receiver and 1.4\,GHz using 19-beam receiver. We study the emission shift events of PSR B1859$+$07 using the simultaneously observed multi-frequency data, and revealed the shift events is frequency-independent. A quasi-periodicity of about 150 pulses was found by analyzing the event interval duration distribution. We discovered a new type of emission shift events with emission at both the shift and non-shift phases. We also identified the two emission states, the B and Q modes, in the non-shift pulses. We analyzed the polarization properties of the three emission states, and found that the three modes have almost the same PA distribution. We argue that none of the models in the literature explains the phenomena of this pulsar well.


\normalem
\begin{acknowledgements}
This work is supported by the National Key $R \& D$ Program of China under grant number 2018YFA0404703, the Open Project Program of the CAS Key Laboratory of FAST, NAOC, Chinese Academy of Sciences, and the Guizhou Education Department under grant No. Qian Education Contract KY[2019]214. This work is supported by the West Light Foundation of the Chinese Academy of Sciences (No. 2018-XBQNXZ-B-023), the``Tianchi Doctoral Program 2021". This work made use of the data from FAST (Five-hundred-meter Aperture Spherical radio Telescope). FAST is a Chinese national mega-science facility, operated by the National Astronomical Observatories, Chinese Academy of Sciences.
\end{acknowledgements}
  
\bibliographystyle{raa}
\bibliography{bibtex}

\end{document}